\documentclass{jkas}

\def\year{2014} 
\def\month{March} 
\def\volume{00} 
\def\issue{0} 
\def\beginpage{1} 
\def\endpage{99} 
\def\received{---} 
\def\accepted{---} 

\date{Received \received ; accepted \accepted}

\title{
Microlens Masses from 1-D Parallaxes and Heliocentric Proper Motions
}

\author{Andrew Gould}

\affil{Department of Astronomy Ohio State University,
140 W.\ 18th Ave., Columbus, OH 43210, USA 
\email{gould@astronomy.ohio-state.edu }}

\def\corrauthor{
Andrew Gould
}

\def\runningauthor{
Andrew Gould
}

\def\runningtitle{
1-D Microlens Parallaxes
}

\def\keywords{
astrometry -- gravitational microlensing -- planets -- stars: fundamental parameters (masses)
}

\def\abstracttext{
One-dimensional (1-D) microlens parallaxes can be combined with heliocentric
lens-source relative proper motion measurements to derive the lens mass
and distance, as suggested by Ghosh et al.\ (2004).  Here I present
the first mathematical anlysis of this procedure, which I show
can be represented as a quadratic equation.  Hence, it is formally
subject to a two-fold degeneracy.  I show that this degeneracy can
be broken in many cases using the relatively crude 2-D parallax information
that is often available for microlensing events.  I also develop
an explicit formula for the region of parameter space where it is
more difficult to break this degeneracy.  Although no mass/distance
measurements have yet been made using this technique, it is likely
to become quite common over the next decade.
}

\newcommand{\bdv}[1]{\mbox{\boldmath$#1$}}

\def\au{{\rm AU}} 
 
\def\bv{{\bf v}}

\def\kms{{\rm km}\,{\rm s}^{-1}}
\def\masyr{{\rm mas}\,{\rm yr}^{-1}}

\def\rel{{\rm rel}}

\def\hel{{\rm hel}}
\def\geo{{\rm geo}}

\def\e{{\rm E}}
\def\bpi{{\bdv\pi}}
\def\bmu{{\bdv\mu}}

\def\btheta{{\bdv\theta}}

\def\apj{{ApJ}}
\def\apjl{{ApJL}}

\begin{document}
\jkashead 


\section{{Introduction}
\label{sec:intro}}

Even at the dawn of microlensing experiments 20 years ago,
it was already realized that vastly more microlensing
events will yield one-dimensional (1-D) microlens parallaxes than full
(2-D) parallaxes \citep{gmb94}.  However, these 1-D parallaxes remain largely 
unexploited.

The amplitude of the microlens parallax vector $\bpi_\e$ is the
ratio of the lens-source relative parallax $\pi_\rel$ to the
angular Einstein radius $\theta_\e$, while its direction is set by the
lens-source relative proper motion $\bmu$,
\begin{equation}
\bpi_\e = {\pi_\rel\over\theta_\e}\,{\bmu\over\mu}.
\label{eqn:bpie}
\end{equation}
The numerator of the first ratio quantifies the apparent angular
lens-source displacement as the observer changes position, while
the denominator translates this angular scale into the scale
of microlensing phenomena.  The second term appears because the
evolution of microlensing effects due to parallax depends on the
direction of lens-source relative motion.  See Figure 1 of
\citet{gouldhorne}.

Note that from the definition of $\theta_\e$,
\begin{equation}
\theta_\e^2 \equiv \kappa M \pi_\rel ,
\qquad \kappa \equiv {4G\over c^2\au}\simeq 8.1{{\rm mas}\over M_\odot} ,
\label{eqn:thetae}
\end{equation}
measurement of both $\theta_\e$ and $\pi_\e$ directly leads to determinations
of the lens mass and relative parallax
\begin{equation}
M = {\theta_\e\over \kappa \pi_\e},
\qquad
\pi_\rel = \theta_\e\pi_\e .
\label{eqn:mpirel}
\end{equation}

The reason that 1-D parallaxes are easier to measure is that, to
the extent that $\bmu$ is parallel to 
the direction of Earth's acceleration (projected
on the sky at the peak of the event) ${\bf \hat n}_a$,
it causes the event evolution to speed up or slow down
as it progresses, thereby inducing an asymmetric distortion on the
lightcurve.  Since microlensing lightcurves are otherwise intrinsically
symmetric, such distortions are easily measured.  By contrast,
to the extent that $\bmu$ is perpendicular to ${\bf \hat n}_a$, it
gives rise to a symmetric distortion, which is much more difficult
to disentangle from other symmetric parameters.  Hence, we define
\begin{equation}
\pi_{\e,\parallel} \equiv \bpi_{\e,\geo}\cdot {\bf \hat n}_a
\label{eqn:piepar}
\end{equation}
and $\pi_{\e,\perp} = (\pi_\e^2 - \pi_{\e,\parallel}^2)^{1/2}$ (see \citealt{gould04}
for sign conventions).

A very important notational point for the current work is that
$\bpi_\e$ is subscripted with ``geo'' (to indicate geocentric frame
at the peak of the event) whereas $\pi_\e$ is not.  This is because
the magnitude of this vector does not depend on the frame of reference,
but the direction does (due to explicit dependence on $\bmu/\mu$
in Equation~(\ref{eqn:bpie})).

Despite their predicted ubiquity, 1-D parallaxes are almost never reported
in the literature.  Among the handful of exceptions 
\citep{park04,ghosh04,jiang04,ob05071b,mb09266}, the first three
reports are due to the novelty of the phenomenon and the last two
are due to the fact that the actual value of $\pi_{\e,\parallel}$
was so large that $\pi_\e$ (and so $M$) could be reasonably estimated
despite the poor measurement of $\pi_{\e,\perp}$.

Nevertheless, as first pointed out by \citet{ghosh04}, 1-D microlens
parallaxes could yield microlens mass (and distance) measurements
simply by making a late-time measurement of the heliocentric
proper motion.  Stated in the most naive way, knowing one component
($\pi_{\e,\parallel}$) of the vector $\bpi_\e$
(from the microlensing event) and
its direction $\bmu/\mu$ (from late-time astrometry), one can
extract the amplitude of the full vector $\pi_\e$.  Then
knowing the Einstein timescale $t_\e$ (again from the event) and
the amplitude of proper motion $\mu$, one can determine the
angular Einstein radius $\theta_\e = \mu t_\e$.

The problem with this naive reasoning, as already recognized by
\citet{ghosh04}, is that $\pi_{\e,\parallel}$ and $t_\e$ are derived
in the geocentric frame, whereas $\bmu$ is measured in the heliocentric
frame.  The relation between these frames, as noted by \citet{ghosh04}
but in a form closer to that preferred by \citet{mb08310}, is
\begin{equation}
\bmu_\hel = \bmu_\geo + \bmu_\oplus\pi_\rel
= {\theta_\e\over t_\e}{\bpi_{\e,\geo}\over \pi_\e} + \bmu_\oplus\pi_\e\theta_\e,
\label{eqn:bmuhel}
\end{equation}
where 
$\bmu_\oplus\equiv \bv_{\oplus,\perp}/\au$ and $\bv_{\oplus,\perp}$ is the
transverse velocity of Earth in the frame of the Sun at the
peak of the event, projected on the plane of the sky.

However, in the intervening 10 years, there has been essentially no work aimed
at understanding Equation~(\ref{eqn:bmuhel}).  For example, it has not even been
recognized that this equation leads to a quadratic
equation in $\pi_{\e,\perp}$, which therefore has two (or zero)
real solutions.

Nevertheless, many changes in the observational landscape are leading
to radical improvements for the prospects of implementing the
original suggestion of \citet{ghosh04}.  First is the simple
fact that the massive search for events made possible by the
inauguration of the Optical Gravitational Lensing
Experiment (OGLE-III) survey in 2002 are now more than 10 years
old, implying that sources and lenses have already substantially
separated.  Second, the Giant Magellan Telescope (GMT) is
already under construction and likely to be operational within
another 10 years.  When it is, its FWHM in $J$ band will be
about 11 mas, meaning that sources and lenses separated by just
2 FWHM (22 mas) will be eligible for very good $\bmu_\hel$ measurements.
Third, microlensing event detections have already increased dramatically
since 2011 with the inauguration of the OGLE-IV survey and
are likely to accelerate further in 2015 with the inauguration of 
the new Korea Microlensing Telescope Network (KMTNet) of 1.6m telescopes
with $4\,{\rm deg}^2$ cameras on three continents.  Since
typical lens-source proper motions are $\mu\sim 4\,\masyr$,
many of these events will be accessible to GMT.  Finally, it
is quite plausible that new space-based wide-field survey telescopes,
like {\it Euclid} and {\it WFIRST} will survey essentially the entire 
bulge microlensing
field at roughly the same epoch as GMT first light.  Since these
have much smaller apertures than GMT, their FWHM will be much larger
(roughly 140 mas for {\it Euclid} and 110 for {\it WFIRST}).  However,
their greater point-spread-function stability will permit 
proper-motion measurements at 1 FWHM (rather than 2), and their
wider (i.e., systematic) coverage will permit a systematic search
for relatively high proper-motion archival events over a very
large field.

Of course, once such data are obtained, the analysis will proceed
by simultaneous fitting of the microlensing and astrometric data.
However, a proper analytic investigation of Equation~(\ref{eqn:bmuhel})
is still very important for understanding what can be learned
from such observations, which is the central motivation for taking
them.

\section{{Quadratic Form}
\label{sec:quadratic}}

Simply taking the ratio of the two components (perpendicular and parallel
to ${\bf \hat n}_a$) of $\bmu_\hel$ (Equation~(\ref{eqn:bmuhel}))
yields the tangent of the angle $\phi$
of this vector relative to ${\bf \hat n}_a$,
\begin{equation}
\tan\phi \equiv {\mu_{\hel,\perp}\over\mu_{\hel,\parallel}}
= {\pi_{\e,\perp} + \theta_{\oplus,\perp} \pi_\e^2
\over \pi_{\e,\parallel} + \theta_{\oplus,\parallel} \pi_\e^2};
\quad \btheta_\oplus\equiv \bmu_\oplus t_\e.
\label{eqn:phidef}
\end{equation}
Note that
$\theta_\oplus = 0.23(v_{\oplus,\perp}/20\,\kms)(t_\e/20\,{\rm day})$.
Then writing $\pi_\e^2 = \pi_{\e,\parallel}^2 + \pi_{\e,\perp}^2$ and
rearranging terms yields
\begin{equation}
A\pi_{\e,\perp}^2 - \pi_{\e,\perp} + C = 0,
\label{eqn:quad}
\end{equation}
where
\begin{equation}
A \equiv \theta_{\oplus,\parallel}\tan\phi-\theta_{\oplus,\perp};
\quad
C \equiv A\pi_{\e,\parallel}^2 + \pi_{\e,\parallel}\tan\phi,
\label{eqn:AandC}
\end{equation}
which has the solutions
\begin{equation}
\pi_{\e,\perp} = {1\pm\sqrt{1 - 4 A C}\over 2 A}.
\label{eqn:quadeq}
\end{equation}

\subsection{{Limit of $4AC\ll 1$}
\label{sec:limit}}

In the limit $4AC\ll 1$, the two solutions of Equation~(\ref{eqn:quadeq}) 
can be approximated as
\begin{equation}
\pi_{\e,\perp} = C(1  + AC + \ldots)
\qquad (4AC\ll 1)
\label{eqn:smallAC}
\end{equation}
and
\begin{equation}
\pi_{\e,\perp} =  {1\over A} - C + \ldots
\qquad (4AC\ll 1,\quad {\rm alternate}).
\label{eqn:smallAC2}
\end{equation}
Note that both solutions are perfectly valid.  The leading term
in the first solution corresponds to ignoring the first term
in Equation~(\ref{eqn:quad}) and yields ``small'' $\pi_{\e,\perp}$, while 
the leading term in the second solution corresponds to ignoring the last term
in Equation~(\ref{eqn:quad}) and yields ``large'' $\pi_{\e,\perp}$.

In practice, there will be some information about $\pi_{\e,\perp}$ from
the microlensing event, which may well be adequate to break this
degeneracy.  The important point is, however, that these two solutions
are likely to correspond to distinct minima, so that they will not
both automatically be probed by simple downhill minimization:
they must both be explicitly checked.  This situation is similar
to the ``jerk-parallax'' degenerate solutions, which also constitute the roots
of quadratic equation in $\pi_{\e,\perp}$.  See
Equation (20) of \citet{gould04}. In fact, prior
to the discovery of this degeneracy, \citet{alcock01} found one of the
two solutions by downhill minimization.  \citet{gould04} then found the
second solution from the symmetries of the quadratic equation, which
turned out to have equally good $\chi^2$.  Subsequent astrometric
measurements by \citet{drake04} showed that the second solution was
correct \citep{gould04b}.

\subsection{{General Case}
\label{sec:gencase}}

Equation~(\ref{eqn:quadeq}) can be reformulated to eliminate ``$C$''
in favor of direct observables
\begin{equation}
\pi_{\e,\perp} = 
{1\pm \sqrt{\sec^2\phi - (2 A\pi_{\e,\parallel} + \tan\phi)^2}\over 2A}.
\label{eqn:general}
\end{equation}
Thus, if $2A\pi_{\e,\parallel}\sim \sec\phi-\tan\phi$, then the argument
under the radical (discriminant) is close to zero and there is a danger that the
error ellipses from the two solutions merge and/or cannot be distinguished
by the microlensing data.  Thus checking for this approximate
equality is an important diagnostic.  

If the source lies on the ecliptic, then $\theta_{\oplus,\perp}=0$.
Because microlensing fields lie close to the ecliptic, this
can often be a useful approximation.  In this case
\begin{equation}
\pi_{\e,\perp} = 
{\cot\phi\pm \sqrt{\csc^2\phi - (1+2\theta_{\oplus,\parallel}
\pi_{\e,\parallel})^2}\over 2\theta_{\oplus,\parallel}};
\quad (\theta_{\oplus,\perp}\equiv 0).
\label{eqn:general2}
\end{equation}
In this form, it is clear that as 
$\phi$ tends toward $\pm\pi/2$,  the discriminant will very likely
be small (and hence the solution prone to degeneracy).

Another key point is that even if $\pi_{\e,\parallel}=0$, it may still
be possible to measure $\pi_{\e,\perp}$ (and so $\pi_\e$) from the
direction of $\bmu_\hel$.  This would not be possible from the naive
perspective outlined in Section~\ref{sec:intro}.  Explicitly,
\begin{equation}
\pi_{\e,\perp} = {1\over A} 
= {1\over \theta_{\oplus,\parallel}\tan\phi - \theta_{\oplus,\perp}}
\qquad (\pi_{\e,\parallel} = 0).
\label{eqn:pieparzero}
\end{equation}
This is important, particularly if there are wide-field high-resolution
survey data (from e.g., {\it Euclid} or {\it WFIRST}) for which special
observations of ``low-probability'' (i.e., $\pi_{\e,\parallel}\sim 0$) 
targets are not
required.  However, events that have high priority (such as those
with planetary events) with $\pi_{\e,\parallel}\sim 0$
could be targeted for individual observations to measure $\bmu_\hel$,
particularly if the planetary-event lightcurve indicated a high 
geocentric scalar proper motion $\mu_\geo$ \citep{henderson14}.

\section{{Error Analysis}
\label{sec:error}}

While $\bv_{\oplus,\perp}$ is known exactly, none of the quantities
entering Equation~(\ref{eqn:quad}) are known exactly.  In particular,
$\btheta_{\oplus}= \bv_{\oplus,\perp}t_\e/\au$, and $t_\e$ is a measured
quantity from the event.  Similarly, $\pi_{\e,\parallel}$ is measured from
the event, while $\tan\phi$ comes from the proper motion measurement.
However, in most cases, $\pi_{\e,\parallel}$ will be measured with
substantially worse precision than the other quantities.  Hence,
it is useful to approximate the others as ``known perfectly'' and
ask how the uncertainty in $\pi_{\e,\perp}$ depends on the measurement
of $\pi_{\e,\parallel}$.  To determine this, I differentiate 
Equation~(\ref{eqn:quad}) and find
\begin{equation}
{\delta \pi_{\e,\perp}\over \delta \pi_{\e,\parallel}}
= {\tan\phi + 2A\pi_{\e,\parallel}\over 1 - 2A\pi_{\e,\perp}}.
\label{eqn:deriv}
\end{equation}
Clearly the appearance of $\pi_{\e,\perp}$ in the denominator of the 
right hand side
reflects a fundamental shortcoming of making a linearized analysis
of an intrinsically non-linear problem.  Nevertheless, this expression
points to the possibility of a strong degeneracy if 
$\pi_{\e,\perp}\sim 1/2A$.  Note from Equation~(\ref{eqn:quadeq}) that this
corresponds to the discriminant $(1-4AC)$ being close to 
zero\footnote{It is a generic property of quadratic equations that
if the coefficients depend on some quantity $q$, then the ratio of
the error in the solution to the error in $q$ diverges as the discriminant
approaches zero}.  
Recall from
Section~\ref{sec:gencase} that this is the same region of solution
space that is potentially most sensitive to the discrete degeneracy.
Therefore, measurements of $\pi_{\e,\parallel}\sim 1/2A$ are particularly
problematic.  Note, however, from comparison with 
Equation~(\ref{eqn:pieparzero}) that this degeneracy specifically
does not apply to the $\pi_{\rm E,\perp}=0$ case.

\section{{Information from Direct Imaging}
\label{sec:photo}}

Of course, if the lens is separately imaged from the source, it is
possible in principle to make a photometric estimate of its mass
and distance.  However, such measurements face a number of challenges.
First, the photometry will most often be done in the near-infrared,
a spectral region for which reddening is highly degenerate with
intrinsic temperature.  Second, even if the lens is in the Galactic
bulge (and so behind essentially all the dust, whose extinction
properties can then be measured from nearby clump stars, e.g.,
\citealt{nataf13}), even main-sequence stars can differ in mass by several
tens of percent at fixed color.  Finally, and most fundamentally,
most stars are in binaries, and for a significant minority of cases
it will be the lower-mass component that gives rise to the microlensing
event (because event rate scales as $M^{1/2}$), 
while the more massive component will be seen by direct imaging
(because light is a high power of mass).  

Note that while in high-resolution
imaging of microlensing lens/source pairs carried out to date it has
often been possible to rule out binary companions over most of
parameter space, this will not be possible in the imaging of more
generic events in the future.  For example, \citet{mb11293B} were
able to rule out companions closer than 11 mas.  This corresponded
to physical separations less than 80 AU.  However, the basis
for this limit was that the event (MOA-2011-BLG-293) had very
high magnification, making it very sensitive to binary companions.
By contrast, typical events with 1-D parallaxes will not be high-magnification,
implying that microlensing-based constraints on companions will be
very weak.

Thus, overall, mass/distance determinations 
from $(\pi_{\e,\parallel},\bmu_\hel)$ will be more accurate than those
derived from photometry, and in a substantial minority of cases
there will be significant disagreement due to lensing by secondary
components of binaries.  It is true that in these cases the
initial $\bmu_\hel$ measurement will be in error due to an implicit
assumption that the primary and the source were initially aligned,
whereas actually the alignment was with the unseen (or barely seen)
secondary.  However, once these cases are identified, they can
be rectified by a second epoch of imaging, which would directly measure
the proper motion of the primary.  This will differ from the
proper motion of the secondary (i.e., the lens) due to
orbital motion, but usually by an amount that is very small
compared to the lens-source proper motion itself.

In brief, photometric mass/distance estimates can be an important
check on estimates derived from a combination of 1-D parallaxes
and heliocentric proper motions, but they will generally be
less precise and less accurate.

\section{{Conclusion}
\label{sec:conclusion}}

Determinations of microlens masses from the combination of 1-D
microlens parallax $(\pi_{\e,\parallel})$ and heliocentric proper motion
($\bmu_\hel$) are likely to become quite common over the next decade.
I have shown that solutions derived from such data are in the form
of a quadratic equation and therefore have an intrinsic two-fold
degeneracy.  This degeneracy may be broken by microlensing data,
which weakly constrain $\pi_{\e,\perp}$ even when they are unable
to measure it precisely.  The degeneracy is most severe when
the discriminant of the quadratic equation is near zero.  In this
case, each solution separately has large errors, so that the two
solutions may merge together.  Photometric mass/distance estimates
can be an important check on mass/distance determinations based
on 1-D parallaxes and heliocentric proper motions, but are overall
less accurate and less precise.



\acknowledgments

I thank Radek Poleski for stimulating discussions.
This work was supported by NSF grant AST 1103471 and NASA grant NNX12AB99G.




\begin{thebibliography}{}


\bibitem[Alcock et al.(2001)]{alcock01}
Alcock, C., Allsman, R.A., Alves, D.R. et al. 2001, 
Direct detection of a microlens in the Milky Way,
Nature, 414, 617

\bibitem[Batista et al.(2014)]{mb11293B} Batista, V., Beaulieu, J.-P.,
Gould, A., Bennett, D.P., Yee, J.C., Fukui, A., Sumi, T., \& Udalski, A.\ 2014,
 MOA-2011-BLG-293Lb: First Microlensing Planet Possibly in the Habitable Zone,
\apj, 780, 54

\bibitem[Dong et al.(2009)]{ob05071b} 
Dong, S., Gould, A., Udalski, A., et al. 2009, 
OGLE-2005-BLG-071Lb, the Most Massive M Dwarf Planetary Companion?,
\apj, 695, 970

\bibitem[Drake et al.(2004)]{drake04}
Drake, A.J., Cook, K.H., \& Keller, S.C. 2004, 
Resolving the Nature of the Large Magellanic Cloud Microlensing Event 
MACHO-LMC-5,
\apj, 607, L29

\bibitem[Ghosh et al.(2004)]{ghosh04}
Ghosh, H., DePoy, D.L., Gal-Yam, A. et al.\ 2004, 
Potential Direct Single-Star Mass Measurement,
\apj, 615, 450

\bibitem[Gould(2000)]{gould00} Gould, A. 2000, 
A Natural Formalism for Microlensing,
\apj, 542, 785

\bibitem[Gould(2004)]{gould04} Gould, A. 2004, 
Resolution of the MACHO-LMC-5 Puzzle: The Jerk-Parallax Microlens Degeneracy,
\apj, 606, 319,

\bibitem[Gould et al.(2004)]{gould04b}
Gould, A, Bennett, D.P., \& Alves, D.R. 2004, 
The Mass of the MACHO-LMC-5 Lens Star,
\apj, 614, 404

\bibitem[Gould \& Horne(2013)]{gouldhorne}  Gould, A. \& Horne, K. 2013, 
Kepler-like Multi-plexing for Mass Production of Microlens Parallaxes
\apjl, 779, L28 

\bibitem[Gould et al.(1994)]{gmb94} Gould, A.,
Miralda-Escud\'e, J. \&  Bahcall, J.N. 1994, 
Microlensing Events: Thin Disk, Thick Disk, or Halo?,
\apj, 423, L105

\bibitem[Henderson et al.(2014)]{henderson14}  Henderson, C.B., Park, H.,
Sumi, T. et al. 2014, 
Candidate Gravitational Microlensing Events for Future Direct Lens Imaging,
\apj, 794, 71 

\bibitem[Janczak et al.(2010)]{mb08310}Janczak, J., Fukui, A., Dong, S.,
 et al. 2010, 
Sub-Saturn Planet MOA-2008-BLG-310Lb: Likely to be in the Galactic Bulge,
\apj, 711, 731

\bibitem[Jiang et al.(2004)]{jiang04}
Jiang, G., DePoy, D.L., Gal-Yam, A. et al.\ 2004, 
OGLE-2003-BLG-238: Microlensing Mass Estimate of an Isolated Star,
\apj, 617, 1307

\bibitem[Muraki et al.(2011)]{mb09266} Muraki, Y., Han, C., Bennett, D.P.,
et al.\ 2011, 
Discovery and Mass Measurements of a Cold, 10 Earth Mass Planet and Its Host 
Star
\apj, 741, 22

\bibitem[Nataf et al.(2013)]{nataf13} Nataf, D.M., Gould, A., 
Fouqu\'e, P. et al. 2013, 
Reddening and Extinction toward the Galactic Bulge from OGLE-III: 
The Inner Milky Way's RV ~ 2.5 Extinction Curve,
\apj, 769, 88

\bibitem[Park et al.(2004)]{park04}
Park B.-G., DePoy, D.L., Gaudi, B.S. et al.\ 2004, 
MOA 2003-BLG-37: A Bulge Jerk-Parallax Microlens Degeneracy,
\apj, 609, 166


\end{thebibliography}
\end{document}